\documentclass[aip,reprint]{revtex4-1}
\usepackage{dcolumn}
\usepackage{bm}
\usepackage[mathlines]{lineno}

\usepackage{amsmath}
\usepackage{amsfonts}
\usepackage{mathtools}
\usepackage[utf8]{inputenc}
\usepackage[T1]{fontenc}
\usepackage{mathptmx}
\usepackage{xcolor}
\usepackage{graphicx}

\newcommand{\ee}[1]{\mathrm{e}^{#1}}
\begin{document}


\title{Single-file diffusion in a bi-stable potential: signatures of memory in the barrier-crossing of a tagged-particle} 



\author{Alessio Lapolla}

\author{Alja\v{z} Godec}
\email[]{agodec@mpibpc.mpg.de}
\affiliation{Mathematical bioPhysics Group, Max Planck Institute for
  Biophysical Chemistry, Am Fassberg 11, 37077 G\"{o}ttingen, Germany}


\begin{abstract}
We investigate memory effects in barrier-crossing in the overdamped
setting. We focus on the scenario where the hidden degrees of freedom
 relax on exactly the same time scale as the
observable. As a prototypical model we analyze
tagged-particle diffusion in a single-file confined to a bi-stable potential. 
We identify the signatures of memory and explain their origin.
The emerging memory is a result of the projection of collective many-body
eigenmodes onto the motion of a tagged-particle. We are interested
in the 'confining' (all background particles in front of the
tagged-particle) and  'pushing' (all background particles behind the
tagged-particle) scenarios for which we find non-trivial and qualitatively different relaxation
behavior. Notably and somewhat unexpectedly, at fixed particle number
we find that the higher the barrier the \textcolor{black}{stronger are
 memory effects}. \textcolor{black}{The fact that the external potential
alters the memory is important more generally, and should be taken
into account in applications of generalized Langevin
equations.}
Our results can readily be tested experimentally and may be
relevant for understanding transport in biological ion-channels.
\end{abstract}

\pacs{}

\maketitle 
\section{Introduction}
Non-linear stochastic flows are at the heart of thermally driven
processes in systems whose potential energy surfaces are characterized
by multiple local energy minima. Pioneered by the seminal work of Kramers
\cite{kramers_brownian_1940} the concept of thermally activated
barrier-crossing has ever since been applied to diverse phenomena,
incl. chemical reactions \cite{Nitzan,Schlogl_1972,Matheson}, tunnel-diodes \cite{Landauer}, laser-pumping \cite{Risken_1965},
magnetic resonance \cite{Kubo}, conformational dynamics and folding of
proteins \cite{lannon_force-clamp_2013, chung_protein_2018,
  yu_energy_2012, manuel_reconstructing_2015, truex_testing_2015,
  chung_single-molecule_2012, de_sancho_molecular_2014,
  best_diffusive_2006,Netz_2018} and nucleic acids 
\cite{neupane_transition-path_2015, neupane_direct_2016}  and
receptor-ligand binding \cite{rico_heterogeneous_2019}, to name but a few.

From a theoretical point of view, the most detailed and precise results were obtained in the
context of relaxation phenomena \cite{van_Kampen_1977,Caroli_1979,Saito,Bunde,Hanggi_bistable,perico_positional_1993,
  perico_torsional_1994} and
first passage time statistics \cite{hanggi_reaction-rate_1990,
  felderhof_escape_2008, berdichevsky_one-dimensional_1996,
  jun_one-dimensional_2002, fleming_activated_1986,
  berezhkovskii_communication_2018,
  hartich_duality_2018,hartich_interlacing} in Markovian (i.e. memory-less)
systems. However, physical observables typically correspond to
lower-dimensional projections and the observed dynamics is Markovian only
under quite restrictive conditions on the nature of the projection \cite{lapolla_manifestations_2019}.
Quoting van Kampen: "Non-Markov is the rule, Markov is the
exception" \cite{van_kampen_remarks_1998}.

Over the years non-Markovian barrier
crossing has therefore received special attention.
Most approaches considered a generalized Langevin
equation in the underdamped regime with diverse phenomenological
memory kernels for the velocity
in the high
\cite{hanggi_thermally_1982, okuyama_generalized_1986,
  okuyama_nonmarkovian_1986} and low
\cite{carmeli_non-markoffian_1982, carmeli_nonmarkovian_1983}
viscosity limit. 
In the case of diffusion in double-well potentials unified solutions have
been obtained \cite{carmeli_nonmarkovian_1984}. 
Seminal results on
non-Markovian effects in the crossing of high energy barriers have been
obtained by Mel’nikov and Meshkov
\cite{melnikov_theory_1986}, and were later extended to low barriers by
Kalmykov, Coffey and Titov \cite{kalmykov_thermally_2006}. Important
results on non-Markovian barrier-crossing have been obtained in the context of condensed-phase dynamics
\cite{grote_stable_1980, chandler_statistical_1978,Chakrabarti}. More recent
studies of memory effects in bi-stable potentials have been carried
out in the context of conformational dynamics of
macromolecules 
\cite{Makarov,Makarov2,Netz_2018,kappler_non-markovian_2019} and the role of
hydrodynamic memory in surmounting energy barriers \cite{Seyler_2020}, while recent applications
involve the interpretation of experiments on the folding of a DNA hairpin
\cite{pyo_memory_2019}.

Quite detailed analytical results have also been
obtained for overdamped  non-Markovian stochastic flows in bi-stable
potentials, in particular for exponentially correlated noise \cite{Hanggi_NM1,Hanggi_NM2,Fox,Doering,Sancho}. 
Characteristic of these studies is that the memory is introduced
phenomenologically and/or the systems typically posses slow and
fast degrees of freedom. Thereby, integrating out of fast degrees of freedom
leads to memory, and time-scales
similar to, or longer than, the correlation time are of interest.

Here, we are interested in the scenario where the background degrees of freedom
(i.e. those that become integrated out) relax on exactly the same time scale as the
observable. In particular, we are interested in the relaxation
dynamics of a tagged-particle in a single-file of Brownian particles confined to a bi-stable
potential, and investigate the r\^ole of the height of the potential barrier.
Projecting out particles' positions introduces memory and strongly
breaks Markovianity \cite{lapolla_manifestations_2019}. The more
particles' coordinates become integrated out the stronger Markovianity
is broken \cite{lapolla_manifestations_2019}. A distinguishing
characteristic of our approach with respect to the existing
literature is, therefore, that we do
not introduce memory phenomenologically via a generalized Langevin
equation. Instead, the memory arises explicitly as a result of projecting out
degrees of freedom in an exactly solvable Markovian many-body
system. This is important because any external
  potential in general 
  also affects the memory in the tagged-particle's dynamics
  \cite{ZwanzigLN,Zwanzig_Book,lapolla_manifestations_2019}. One
  therefore may \emph{not} employ ad hoc memory-kernels
  that are independent of the external potential, except when
  the potential \textcolor{black}{does not act background degrees of
    freedom} and the
  \textcolor{black}{interaction between the} background degrees of
  freedom \textcolor{black}{and} the tagged-particle is \textcolor{black}{harmonic or} negligibly weak. 

Single-file models are generically used to describe strongly correlated,
effectively one-dimensional, systems and processes, e.g. biological
channels \cite{hummer_water_2001}, transport in zeolites
\cite{chou_entropy-driven_1999}, crowding effects in gene regulation
\cite{gene,li_effects_2009}, superionic conductors
\cite{richards_theory_1977}, and strongly correlated one-dimensional
soft matter systems in general \cite{taloni_single_2017,Lutz,Lin,Locatelli}. Over the
past years diverse theoretical studies yielded deep insight about the
anomalous tagged-particle diffusion \cite{harris_diffusion_1965, jepsen_dynamics_1965, lizana_single-file_2008, lizana_diffusion_2009,barkai_theory_2009, barkai_diffusion_2010, leibovich_everlasting_2013,flomenbom_single-file_2008,metzler_ageing_2014} and the emergence and meaning of
memory
\cite{lizana_foundation_2010,lapolla_unfolding_2018,lapolla_manifestations_2019}. Single-file diffusion in potential landscapes has been studied by computer
simulations \cite{Goldt_2014}.

It is well known
  that a tagged-particle's diffusion in any homogeneous overdamped system of identically
interacting particles with excluded mutual passage is asymptotically
subdiffusive, i.e the tagged-particle's mean squared displacement
asymptotically scales as $\langle
x^2\rangle \sim t^{1/2}$ (see e.g. Ref.~\onlinecite{Kollman}).
However, the manner in which crowding/steric obstruction and particle
correlations affect memory in
barrier-crossing, and in particular in the relaxation towards
equilibrium \textcolor{black}{and how such memory can be inferred and
  quantified from measurable physical observables}, has so far remained
elusive. \textcolor{black}{Our results are relevant in the context of
  search processes of proteins on DNA in the presence of
  macromolecular crowding involved in transcription regulation, and on
  a conceptual level for 
  transport in ion-channels. More generally, the methodological framework
presented here does not require an analytical solution of the problem
to be known, and can thus also be applied in the analysis of experiments or computer
  simulations.}

In this work we provide in Sec.~\ref{Theory} an
analytical solution to the problem using the coordinate Bethe
ansatz.  In Sec.~\ref{linear} we analyze \textcolor{black}{the equilibrium correlation
functions and underlying linear memory kernel} as a function of the barrier height and number of
particles in the single-file. Sec.~\ref{fixed} addresses the
relaxation to equilibrium from a fixed, non-equilibrium initial
condition of the tagged-particle in the 'confining' and ''pushing'
scenario, respectively. We conclude with a brief discussion incl.
potential applications and extensions of our results.

\section{Theory}\label{Theory}
\begin{figure}
    \centering
    \includegraphics[width=0.48\textwidth]{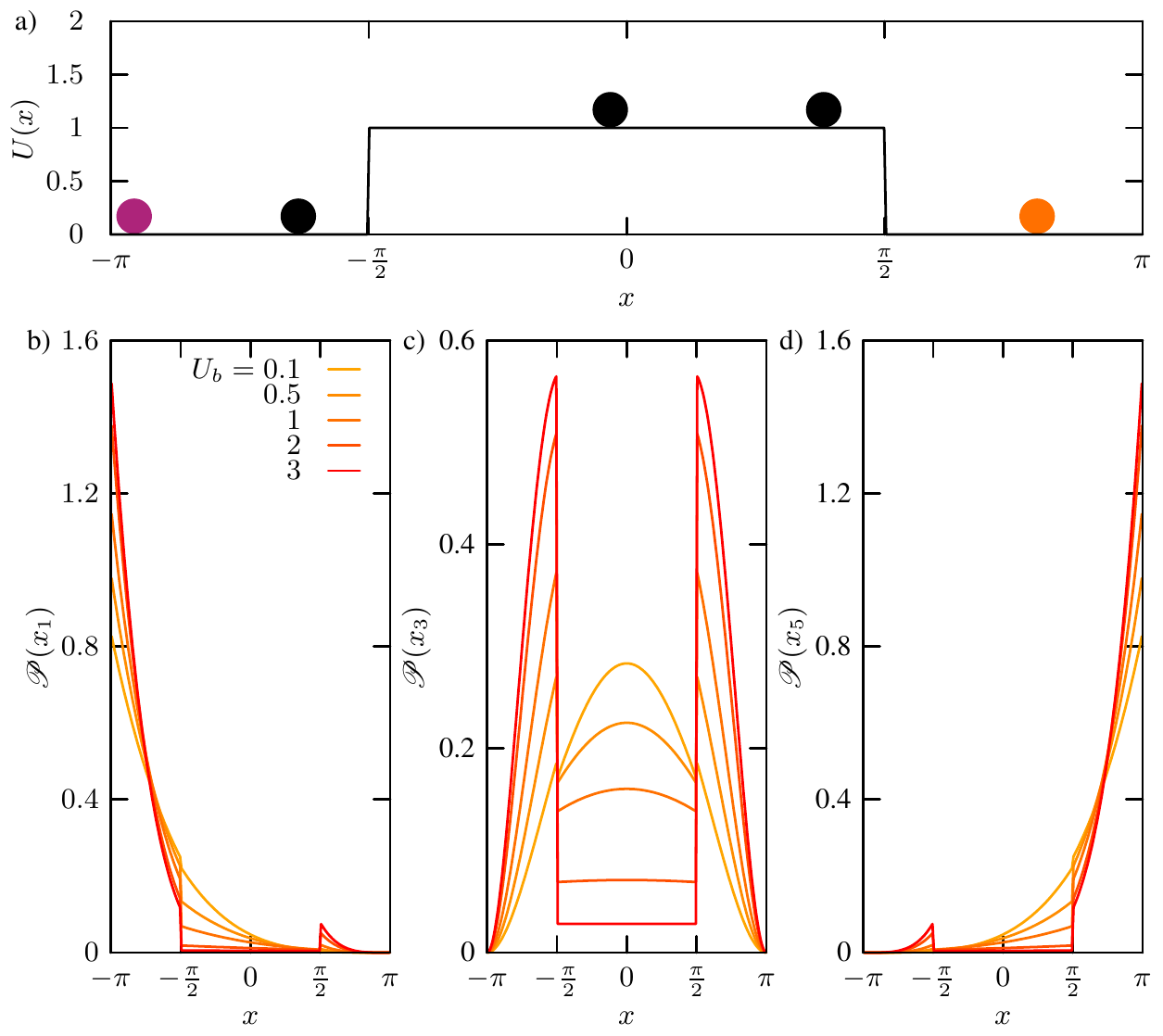}
    \caption{a) Schematic of the potential $U(x)$ defined in
      Eq.~(\ref{pot}); Single-file with  $N=5$
      particles. Throughout this work we either tag the first (magenta
      stands for $i=1$) or the last (orange stands for $i=N$) particle. b),
      c) and d) depict, respectively, the equilibrium probability
      distribution $\mathcal{P}_{\rm eq}(x_i)$ for $i=1$ (b) $i=3$ (c)
      and $i=5$ (d) in a single-file with $N=5$ for different barrier
      heights $U_b$.}
    \label{fig:meanbarrier}
\end{figure}

We consider a single-file of $N$ point-particles confined to a box
of length $L=2\pi$. In the center of the box there is a square-top
energy barrier of width $\pi$ and height $U_b$ (see Fig.~\ref{fig:meanbarrier}a). More precisely, each
particle experiences the potential \cite{morsch_one-dimensional_1979, risken_fokker-planck_1996}
\begin{equation}
U(x)=\begin{cases}
    0, & \pi> |x| > \pi/2\\
    U_b, & |x| \leq \pi/2\\
    \infty, &\text{otherwise}.
\end{cases}
\label{pot}
\end{equation}
The particles move
according to overdamped Brownian dynamics \textcolor{black}{but are not
allowed to cross}. For simplicity and without loss of generality we set
$D=1$, which is equivalent to expressing time in units of
$4\pi^2/D$, and express $U$ in units of thermal energy $k_{\rm B}T$,
i.e. $U\to U/k_{\rm B}T$. The probability density of the set of
positions $\{x_i\}=\mathbf{x}$ of the $N$ particles evolves according to the
many-body Fokker-Planck equation
\begin{equation}
\left(\partial_t-\sum_{i=1}^N\left[
  \partial_{x_i}^2+\partial_{x_i}\{\partial_{x_i}U(x_i)\}\right]\right)G(\mathbf{x},t|\mathbf{x}_0)=0,
\label{FPE}
\end{equation}
with initial condition
$G(\mathbf{x},0|\mathbf{x}_0)=\prod_{i=1}^N\delta(x_i-x_{0i})$ and
where the operator in curly brackets $\{\}$ acts only within the
bracket. Eq.~(\ref{FPE}) is equipped with the set of external and
internal boundary conditions
\begin{eqnarray}
  \partial_{x_1}G(\mathbf{x},t|\mathbf{x}_0)|_{x_1=-\pi}=\partial_{x_N}G(\mathbf{x},t|\mathbf{x}_0)|_{x_N=\pi}&=&0\nonumber\\
  \left(\partial_{x_{i+1}}-\partial_{x_i}\right)G(\mathbf{x},t|\mathbf{x}_0)|_{x_{i+1}=x_i}&=&0,
\label{BC}
\end{eqnarray}
and is solved exactly using the coordinate Bethe ansatz (for technical
details refer to
Refs.~\onlinecite{lapolla_unfolding_2018,lapolla_manifestations_2019,lapolla_bethesf_2020}).
\textcolor{black}{In a nutshell, the Bethe ansatz solution exploits the intuitive
  fact that a trajectory of $N$ identical non-crossing Brownian particles is
  \emph{identical} to that of an ideal Brownian gas if we
  re-label the particle indices such that that they are
  ordered at all times. As a result, one can construct the probability density of the set of
particles' positions $\mathbf{x}$ by a suitable permutation of the products of
probability densities of individual, non-interacting particles. In
turn, an eigenfunction expansion of the many-body Fokker-Planck
operator can be obtained by permuting products of single-particle
eigenspectra, which is what we exploit in the present article.}
The
resulting \textcolor{black}{Bethe} many-body Green's function reads
\begin{equation}
   G(\mathbf{x},t|\mathbf{x}_0)=\sum_{\mathbf{k}} \Psi_\mathbf{k}^R(\mathbf{x})\Psi_\mathbf{k}^L(\mathbf{x}_0)\mathrm{e}^{-\Lambda_\mathbf{k}\tau}
\label{Greens}
\end{equation}
where $\Psi_\mathbf{k}^L(\mathbf{x})$ and
$\Psi_\mathbf{k}^R(\mathbf{x})$ are the so-called left and right Bethe
eigenfunctions, respectively, defined as
\begin{equation}
\Psi_\mathbf{k}^{L,R}(\mathbf{x})\equiv\mathcal{N}^{1/2}\hat{O}_{\mathbf{x}}\sum_{\{\mathbf{k}\}} \prod_{i=1}^N \psi_{k_i}^{L,R}(x_i),
\label{eigf}  
\end{equation}
where $\psi_{n}^{L,R}(x)$ are the orthonormal eigenfunctions of the
single-particle problem (given in Appendix), the sum over
$\{\mathbf{k}\}$ refers to the sum over all permutations of the
multiset $\mathbf{k}$, and $\mathcal{N}$ is the number of these
permutations $\mathbf{k}$. $\Lambda_\mathbf{k}=\sum_{i=1}^N
\lambda_{k_i}$ refers to the Bethe
eigenvalue with multi-index $\mathbf{k}=\{k_i\},i\in [1,N]$, and
$\hat{O}_\mathbf{x}$ is the particle-ordering operator, which ensures
that $x_1\le \cdots\le x_i\le \ldots\le x_N$.
Moreover, $\lambda_{n}$ refer to the eigenvalues of the respective one-body
problem given by \cite{morsch_one-dimensional_1979,risken_fokker-planck_1996}
\begin{equation}
\lambda_n=
    \begin{dcases}
    \frac{n^2}{4},& \mathrm{mod}(n,4)=0,\\
    \left(\frac{n-1}{2}+\nu\right)^2,&  \mathrm{mod}(n,4)=1,\\
    \frac{n^2}{4},& \mathrm{mod}(n,4)=2,\\
    \left(\frac{n+1}{2}-\nu\right)^2,& \mathrm{mod}(n,4)=3
    \end{dcases}
    \label{evs}
\end{equation}
where $\nu=2\arctan(\ee{-U_b/2})/\pi$ and $\mathrm{mod}(k,l)$ stands
for the remainder of the division $k/l$.

We are interested in the
non-Markovian probability density of $x_i$, the position of the $i$-th
tagged-particle under the condition that the initial positions of the
remaining particles are drawn from those equilibrium configurations
that contain particle $i$ at $x_0$, which reads
(for a derivation see
Refs.~\onlinecite{lapolla_unfolding_2018,lapolla_manifestations_2019,lapolla_bethesf_2020})
\begin{equation}
  \mathcal{G}(x_i,t|x_{0i})=V_{\mathbf{0}\mathbf{0}}^{-1}(x_{0i})\sum_\mathbf{k} V_{\mathbf{0}\mathbf{k}}(x_i)V_{\mathbf{k}\mathbf{0}}(x_{0i})\mathrm{e}^{-\Lambda_\mathbf{k}t},
\label{NMGreen}
\end{equation}
where the 'overlap-elements' $V_{\mathbf{k}\mathbf{l}}(x_i)$ are defined as
\cite{lapolla_bethesf_2020}
\begin{equation}
 V_{\mathbf{k}\mathbf{l}}(x_i)=\frac{m_{\mathbf{l}}}{N_L!N_R!}\sum_{\{\mathbf{k}\}}\sum_{\{\mathbf{l}\}}\psi_{k_i}^R(x_i)\psi_{l_i}^L(x_i)
  \prod_{n=1}^{i-1}L_n(x_i)\!\!\!\!\!\prod_{m=i+1}^{N}\!\!\!\!R_m(x_i)
\label{overlap}
\end{equation}  
with $m_{\mathbf{l}}$ being the multiplicity of the multiset
$\mathbf{l}$, and $N_L=i-1$ and $N_R=N-i$ are, respectively, the number of particles to the left
and right of the tagged-particle. In Eq.~(\ref{overlap}) we introduced the auxiliary functions
\begin{equation}
     L_n(x)=\int_{-\pi}^x dz \psi_{l_n}^L(z)\psi_{k_n}^R(z),\quad R_n(x)=\int_{x}^\pi dz\psi_{l_n}^L(z)\psi_{k_n}^R(z). 
\label{aux}  
\end{equation}
Note that the equilibrium probability density of the tagged-particle's
position is given by (see Eq.~(\ref{NMGreen})) $\mathcal{P}_{\rm
  eq}(x_i)\equiv\lim_{t\to\infty}\mathcal{G}(x_i,t|x_{0i})=V_{\mathbf{0}\mathbf{0}}(x_i)$
and is depicted for various values of $U_b$ in
Fig.~\ref{fig:meanbarrier}b-d. Intuitively, as $U_b$ increases particles become
expelled from the barrier.

In Ref.~\onlinecite{lapolla_bethesf_2020} we have developed an
algorithm designed to efficiently cope with the combinatorial
complexity of the implementation of the analytical solution in
Eq.~(\ref{NMGreen}). Due to the piece-wise constant nature of the
potential $U(x)$ in Eq.~(\ref{pot}) all integrals (\ref{aux}) can be
computed analytically. As the resulting expressions are lengthy we do not show them
here. Instead, they are readily implemented in an extension of the
code published in Ref.~\onlinecite{lapolla_bethesf_2020} (see
Supplementary Material). 

 \section{Linear correlations at equilibrium}
 \label{linear}
\begin{figure}
    \centering
    \includegraphics[width=0.48\textwidth]{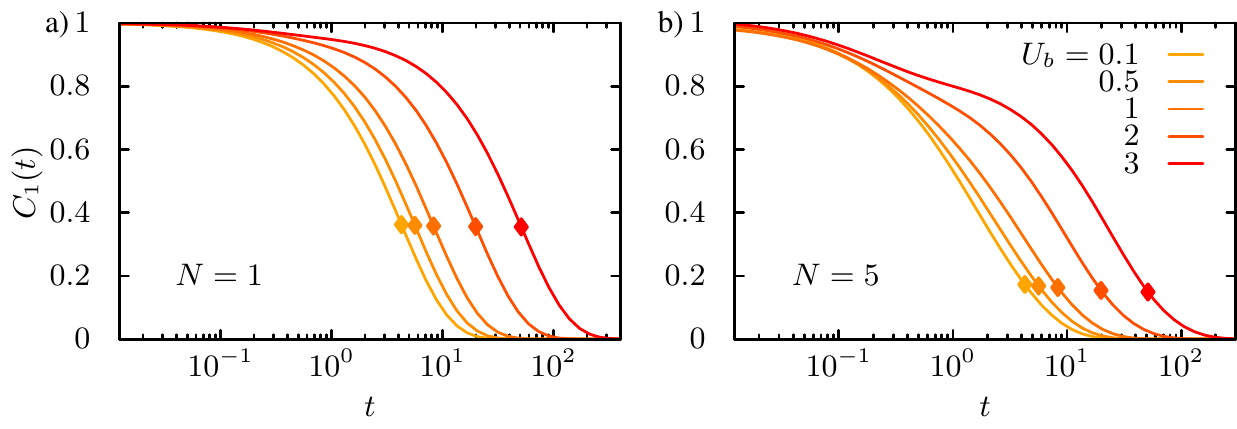}
    \caption{Position autocorrelation function $C_1(t)$ of an isolated particle
      (a) and the leftmost tagged-particle in a single-file with five particles (b)
      as a function of the barrier height $U_b$. Symbols denote $C_1(\Lambda_{\mathbf{1}}^{-1})$.}
    \label{fig:autocorr}
\end{figure}

First we consider linear correlations at equilibrium and limit the
discussion in the reminder of the paper to tagging the first or last particle
(i.e. throughout we set $i=1$ or $i=N$). That is, we are
interested in the normalized positional autocorrelation function of
a tagged-particle defined as
\begin{equation}
    C_i(t)=\frac{\langle x_i(t)x_i(0)\rangle -\langle
    x_i\rangle^2}{\langle x_i^2\rangle -\langle x_i\rangle^2},
\label{autocor}
\end{equation}
where the covariance of the position is defined as
\begin{equation}
    \langle x_i(t)x_i(0)\rangle\equiv\int_{-\pi}^\pi dx_i
    \int_{-\pi}^\pi dx_{0i}\; x_i x_{0i} \mathcal{G}(x_i,t|x_{0i})
    \mathcal{P}_{\rm eq}(x_{0i}),
\end{equation}
and $\langle x_i^n \rangle =\int_{-\pi}^\pi dx_i\; x_i^n
\mathcal{P}_{\rm eq}(x_i)$.
The above integrals have been performed numerically by means of
Gauss-Kronrod quadrature \cite{boost_quadrature}. Note that Eq.~(\ref{autocor}) alongside
Eqs.~(\ref{eigf}-\ref{aux}) necessarily
implies the structure
$C_i(t)=\sum_{\mathbf{k}\ne\mathbf{0}}a_{\mathbf{k}}\mathrm{e}^{-\Lambda_{\mathbf{k}}t}$
with $\sum_{\mathbf{k}\ne\mathbf{0}}a_{\mathbf{k}}=1$ and where all 
$a_{\mathbf{k}}\ge 0$ \footnote{Note that Eqs.~(\ref{eigf}-\ref{aux}) imply
  that $a_{\mathbf{k}}$ is a positive constant times the square of a
  real number.}.
The results for $C_1(t)$ as a function of
the barrier height $U_b$ are depicted in Fig.~\ref{fig:autocorr}. 
Since $U(x)$ is symmetric the autocorrelation functions of the first
and last particle coincide, i.e. $C_1(t)=C_N(t)$.

The autocorrelation of an isolated particle (i.e. $N=1$) in
Fig.~\ref{fig:autocorr}a displays for
a given value of $U_b$ to a good approximation an exponential decay
with rate $\Lambda_{\mathbf{1}}=\lambda_{1}$ given by Eq.~(\ref{evs}). This reflects that
positional correlations decay predominantly due to barrier-crossing.
Conversely, as the
number of particles increases  $C_1(t)$
decays on multiple time-scales (see Fig.~\ref{fig:autocorr}b) and
develops an ``anomalous'' shoulder on shorter time-scales\cite{lizana_foundation_2010}, whose span increases
with the barrier height $U_b$.  A comparison of $C_1(\Lambda_1^{-1})$
reveals that the relative decay of correlations from the
relaxation time $\tau_{\rm rel}\equiv\Lambda_1^{-1}$ onward is
substantially reduced for about a factor of $2$ compared to the
isolated particle case. $\tau_{\rm rel}$ denotes the time-scale on
which the system reaches equilibrium from any initial condition.
Note that (i) $C_1(t)$ measures
relative correlations and (ii) according to Eq.~(\ref{NMGreen})
(terminal) relaxation roughly corresponds to the
particles individually crossing the barrier several times. 
It is also important to note that the
natural time-scale of a tagged-particle is set by the average
collision time\cite{lapolla_unfolding_2018,lapolla_manifestations_2019} $\tau_{\rm col}=1/N^2$  which
decreases with increasing $N$. That is, in units of the average number
of collisions $t\to t/\tau_{\rm col}$ correlations decay more slowly
for larger $N$. 

\begin{figure}
    \centering
    \includegraphics[width=0.48\textwidth]{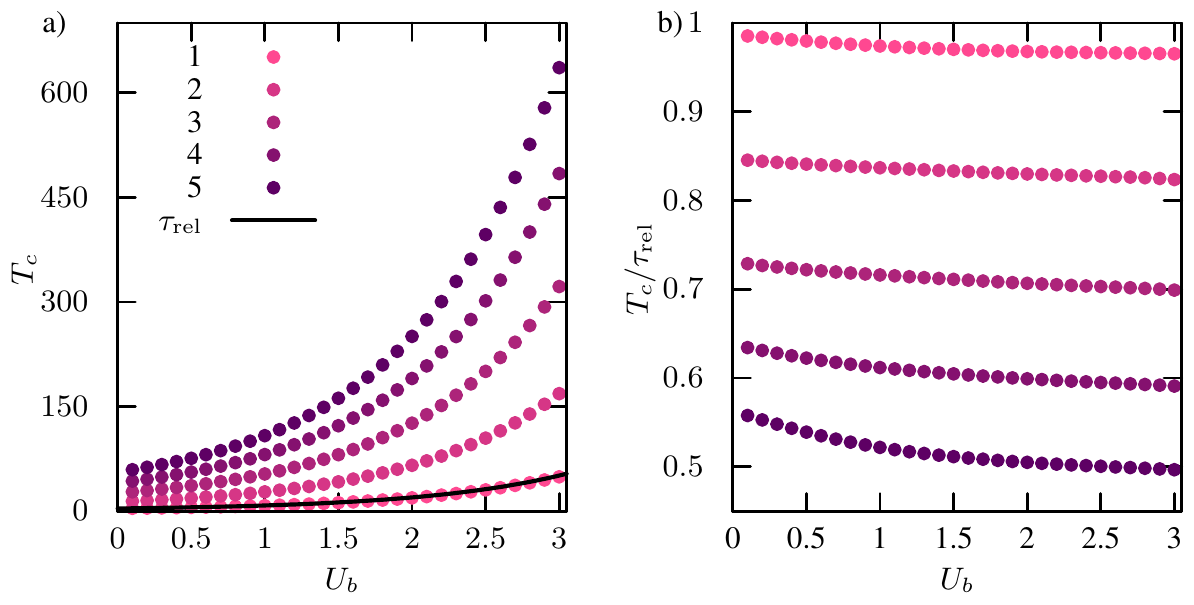}
    \caption{a) $T_c/\tau_{\rm col}$ for the first particle (i.e. $i=1$) as a function of the
      barrier-height $U_b$ for various values of $N$; The full line
      depicts $\tau_{\rm rel}\equiv\Lambda_1^{-1}$. b) Ratio
      $T_c/\tau_{\rm rel}$ as a function of the
      barrier-height $U_b$ for various values of $N$.}
    \label{fig:corrtime}
\end{figure}

A common means to quantify the extent of correlations found in the
literature is the so-called \emph{correlation time} $T_c$ \cite{lipari_model-free_1982,
  perico_positional_1993, perico_torsional_1994,
  kalmykov_thermally_2006} and should be compared with the actual
relaxation time $\tau_{\rm rel}$:\footnote{Note that $\tau_{\rm rel}$
  does not depend on $N$.} 
\begin{equation}
    T_c=\int_0^\infty dt C_i(t), \,\, \tau_{\rm rel}\equiv\Lambda_1^{-1}=\left(\frac{2}{\pi}\arctan(\mathrm{e}^{-U_b/2})\right)^{-2},
\label{ctime}
\end{equation}
where we note that for high barriers, i.e. $U_B\gg 1$, the relaxation time
follows the expected
Arrhenius scaling $\tau_{\rm rel}\simeq 4\mathrm{e}^{U_b}/\pi^2$. In
Fig.~\ref{fig:corrtime}a we depict the correlation time for the
leftmost particle in units of
$\tau_{\rm col}$ as a function of the barrier height $U_b$ for
different $N$.  For an isolated particle $T_c=T_c^{\rm isolated}$ agrees very well with
$\tau_{\rm rel}$ for all values of $U_b$, confirming the idea that
$C(t)$ decays to a very good approximation as a single exponential. Note that, for systems obeying
detailed balance, the mathematical
structure of $C_i(t)$ trivially implies a shorter correlation time as
soon as $C_i(t)$ decays on multiple time-scales if the longest
time-scale $\Lambda_{\mathbf{1}}^{-1}$ is the same. This is particularly
true when comparing $C_i(t)$ of a
tagged-particle in a single-file with an isolated
particle. Namely, 
\begin{equation}
T_c=\sum_{\mathbf{k}\ne\mathbf{0}}a_\mathbf{k}/\Lambda_{\mathbf{k}}\le\sum_{\mathbf{k}\ne\mathbf{0}}a_\mathbf{k}/\Lambda_{\mathbf{1}}=
\Lambda_{\mathbf{1}}^{-1}\approx T_c^{\rm isolated}.   
\label{ineq}
\end{equation}
Therefore, the interpretation of $T_c$ should always be made cautiously and in
the particular case of tagged-particle diffusion in a single-file is not
meaningful if we consider $T_c$ on an absolute scale. However, it becomes somewhat more meaningful
on the natural time-scale, i.e. when time is expressed in terms of the
average number of inter-particle collisions (see also Ref. \onlinecite{lapolla_manifestations_2019}).
Inspecting $C_1(t)$ on this natural time scale we find in
Fig.~\ref{fig:corrtime}a that the
tagged-particle on average undergoes more collisions before it
decorrelates for larger values of $N$, and this number increases with increasing $U_b$.

Moreover, as
$N$ increases the space explored by a tagged-particle becomes
progressively more confined \cite{lapolla_manifestations_2019} rendering the correlation time $T_c$ on an absolute
time-scale also intuitively shorter. Indeed, in Fig.~\ref{fig:corrtime}b we
depict the ratio $T_c/\Lambda_1^{-1}$ which decreases with increasing
$N$ for any barrier
height $U_b$. Note that $\Lambda_1^{-1}$ is independent of $N$ and the
breaking of Markovianity (reflected, e.g. in the violation of the
Chapman-Kolmogorov semi-group property \cite{lapolla_manifestations_2019}) is
encoded entirely in the overlap elements $V_{\mathbf{0k}},V_{\mathbf{k0}}$. For
systems with microscopically reversible dynamics
$T_c/\Lambda_1^{-1}<1$ quite generally implies that relaxation evolves
on multiple time-scales. Thus, the results in Fig.~\ref{fig:corrtime}b
suggest, in agreement with intuition, that more and more time-scales are involved in the relaxation
of a tagged-particle's position in equilibrium as we increase $N$. In
other words, on the level of linear correlations signatures of memory of the initial conditions of
'latent'/background particles are reflected in the multi-scale
relaxation of $C_1(t)$.

As shown by Zwanzig
  from first principles \cite{ZwanzigLN,Zwanzig_Book} one can also analyze the memory
  encoded in $C_i(t)$ defined in Eq.~(\ref{autocor}) in terms of a \emph{memory function} $K_U(t)$ defined through 
\begin{equation}
\frac{d}{dt}C_i(t)=-\int_0^tK_U(s)C_i(t-s)ds,
\label{kernel}  
\end{equation}
where the subscript $U$ is included to stress that the memory kernel
depends on the external potential. \textcolor{black}{The kinetic equation}
(\ref{kernel}) is obeyed \emph{exactly}
\cite{ZwanzigLN,Zwanzig_Book}.

\textcolor{black}{Note that the memory
  kernel $K_U(t)$ in the
  \emph{linear} kinetic equation (\ref{kernel}) is
  \emph{not} equivalent to the memory kernel entering a
  \emph{non-linear} generalized Langevin equation for a
  tagged-particle motion in a potential of mean force
  \cite{Zwanzig_Book,Makarov,Makarov2}. If, however, one were to
  compute $C_i(t)$ from such a non-linear generalized Langevin
  equation this would yield Eq.~(\ref{kernel}). Here, we aim to
  connect quantitatively the different signatures of memory encoded in $C_i(t)$ and
  the correlation time $T_c$ solely by means of the information
  encoded in $C_i(t)$. Note that this approach is simple and
  model-free, and can therefore directly be used in the
  analysis of experimental and simulation data. Alternatively one may
  equally well use a non-linear framework (for an excellent recent
  example see Ref.~\onlinecite{Sollich}) that,
  however, requires more effort as well as a more detailed input.}

We determine $K_U(s)$ from the Laplace transform of Eq.~(\ref{kernel})
\begin{equation}
\tilde{K}_U(u)=\frac{1}{\sum_{\mathbf{k}\ne\mathbf{0}}a_{\mathbf{k}}/(\Lambda_{\mathbf{k}}+u)}-u
\label{Lap_kernel}  
\end{equation}
where $\tilde f(u)\equiv \int_0^\infty\mathrm{e}^{-ut}f(t)dt$. The
inverse Laplace transform is in turn determined by numerically
inverting Eq.~(\ref{Lap_kernel}) using the \emph{fixed Talbot method}
\cite{Abate}. 
\begin{figure}
    \centering
    \includegraphics[width=0.48\textwidth]{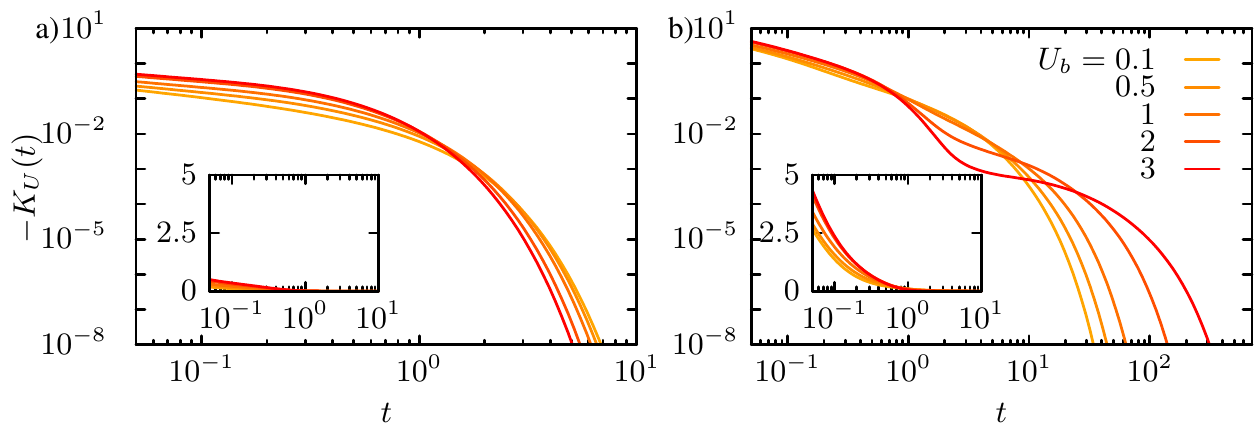}
    \caption{Memory kernel $-K_U(t)$ of an isolated particle
      (a) and the leftmost tagged-particle in a single-file with five particles (b)
      as a function of the barrier height $U_b$ shown on a double
      logarithmic scale. Inset: the same results shown on a
      linear-logarithmic scale. We used 900 pairs
      $a_{\mathbf{k}},\Lambda_{\mathbf{k}}$ to determine
      $\tilde{K}_U(u)$ in Eq.~(\ref{Lap_kernel}). Note that the range
      of $t$ in (a) and (b) is different.}
    \label{memKer}
\end{figure}\\
\indent By construction $\int_0^tK_U(s)C_i(t-s)ds\ge 0$ and therefore $K_U(t)$ must have
a positive contribution at least for $t\to 0$.
For example, if $C_i(t)$ decays
\emph{exactly} as a single exponential with rate $\Lambda$ we have
$\tilde{K}_U(u)=\Lambda$ and hence $K_U(t)=\Lambda \delta(t)$. In fact, one can show by means of
Mori's projection-operator formalism that $K_U(t)$ has the generic
structure $K_U(t)=a\delta(t)+\zeta(t)$ where $a>0$ and $\zeta(t)$ is a
smooth function of $t$ (see e.g. Eq.~(8.61) in Ref.~\onlinecite{Zwanzig_Book}).\\
\indent More generally, when memory is short-lived, that is, when
$K(t)$ decays rapidly compared to the time-scale on which $C_i(t)$
changes appreciably, 
we may approximate Eq.~(\ref{kernel}) as
$\frac{d}{dt}C_i(t)\approx-(\int_0^\infty K(s)ds) C_i(t)\equiv-\Lambda
C_i(t)$ and hence the
auto-correlation decays approximately as a single exponential
$C_i(t)\approx \mathrm{e}^{-\Lambda t}$ -- the
system is therefore said to be effectively memoryless (Markovian)
\cite{ZwanzigLN,Zwanzig_Book}. 
\\
\indent The memory kernel of an isolated particle
 and that of the leftmost tagged-particle in a single-file is depicted
in Fig.~\ref{memKer} (note that we depict $-K_U(t)>0$ implying
an anti-persistent motion). The unavoidable truncation of the spectral solution (\ref{NMGreen}) does not allow us to
determine $K_U(t)$ in the limit $t\to 0$. 
In the case of an isolated particle (Fig.~\ref{memKer}a) the memory
is short-lived and decays on a time-scale $t\ll \Lambda_1^{-1}$ much shorter than the
relaxation time and decreases with the barrier height $U_b$. This
agrees with the essentially single-exponential decay
of $C_1(t)$ found in Fig.~\ref{fig:autocorr} and implies that the
dynamics is essentially memoryless \cite{ZwanzigLN,Zwanzig_Book}.\\
\indent Conversely, in the case of a tagged-particle (see Fig.~\ref{memKer}b)  $K_U(t)$ displays
a strikingly different behavior. First, the range where $K_U(t)$
considerably differs
from zero is orders of magnitude longer and increases with increasing
barrier height $U_b$. Second, the anti-persistence
of $K_U(t)$ is much larger than for an isolated particle, and third for
high barriers
$K_U(t)$ develops a shoulder indicating a transiently stalled decay
of memory, presumably due to  ``jamming'' in front of the
barrier prior to crossing. The latter acts as a transient entropic
trap.\\
\indent Importantly, $U_b$ strongly and non-trivially affects the memory in the
tagged-particle's motion (compare with the trivial effect of $U_b$ on
$K(t)$ for an isolated particle in Fig.~\ref{memKer}a). As a result,
any microscopically consistent memory kernel $K_U(t)$ must depend on
the external potential $U(x)$.
 The reason is twofold: (i) the external potential also
acts on the background degrees of freedom and (ii) the coupling of the
background degrees of freedom and the tagged-particle's motion is
strong. In fact, whenever (i) and/or (ii) hold the external
potential gerenally alters the memory. Conclusive evidence that the
potential affects the memory function has been found e.g. by atomistic computer simulations of molecular solutes in water \cite{Netz_PRX}.\\
\indent In contrast to $C_1(t)$ depicted
Fig.~\ref{fig:autocorr}, which on an absolute time-scale decays to zero faster for larger $N$ as
a result of being normalized and having the same relaxation time
$\Lambda_1^{-1}$, the memory kernel $K_U(t)$ clearly displays long-time
memory effects that become more pronounced as $N$ increases. This can
be understood by noticing that $K_U(t)$ in Eq.~(\ref{kernel}) is
unaffected by the normalization. The memory-kernel is thus more
informative than $C_1(t)$ and less ambiguous than correlation times
$T_c$. Moreover, it is not required that $C_i(t)$ is known
analytically in order to apply the analysis.

\section{Relaxation from a pinned configuration}\label{fixed}
We now focus on the 'complete' (i.e including correlations to all orders) relaxation to equilibrium from a \emph{pinned}
configuration. That is, we are interested in those initial
configurations where either the first ($i=1$) or the last ($i=N$)
particle is pinned at $x_0$, while the initial conditions of the
remaining particles are drawn from the corresponding pinned equilibria
(i.e. those equilibrium many-body configurations where the first/last
particle is located at $x_0$). \textcolor{black}{In this non-stationary
setting the analysis of memory kernels seems less sensible, since
these would depend explicitly on time as well as $x_0$.}  

\begin{figure}
    \centering
    \includegraphics[width=0.48\textwidth]{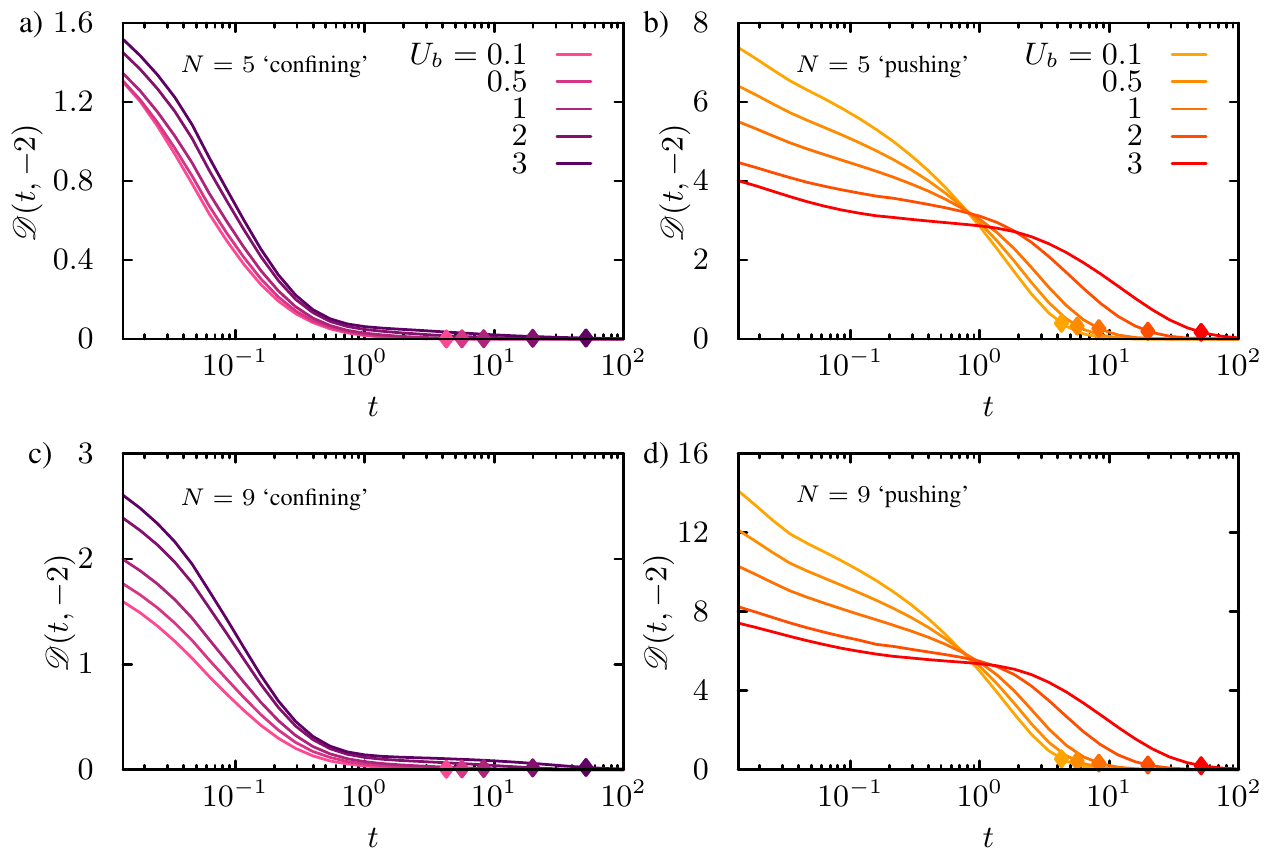}
    \caption{Time evolution of $\mathcal{D}(t,x_{0i})$ for various
      barrier-heights $U_b$ for $N=5$ (a -- confining; b -- pushing)
      and $N=9$ (c -- confining; d -- pushing), respectively. The
      symbols denote $\mathcal{D}(\Lambda_{\mathbf{1}}^{-1},x_{0i})$.}
    \label{fig:klrelax}
\end{figure}

We quantify the relaxation dynamics by means of $\mathcal{D}(t,x_{0i})$, the Kullback-Leibler
divergence \cite{s._kullback_information_1951} between the
non-Markovian probability density of the tagged-particle's position at
time $t$, $\mathcal{G}(x_i,t|x_{0i})$ in Eq.~(\ref{NMGreen}), and the
respective equilibrium density $\mathcal{P}_{\rm
  eq}(x_i)\equiv\lim_{t\to\infty}\mathcal{G}(x_i,t|x_{0i})$:
\begin{equation}
    \mathcal{D}(t,x_{0i})\equiv
    \int_{-\pi}^{\pi} dx
    \mathcal{G}(x,t|x_{0i})\ln\left(\frac{\mathcal{G}(x,t|x_{0i})}{\mathcal{P}_{\rm
      eq}(x)}\right).
    \label{kldiv}
\end{equation}
In
physical terms  $\mathcal{D}(t,x_{0i})$ represents the displacement
from equilibrium in the sense of an
\emph{excess instantaneous free energy}, i.e. $k_{\rm
  B}T\mathcal{D}(t,x_{0i})=F(t)-F$
\cite{Mackey_1989,Qian_2013,Lapolla_PRL}. Since the integral in
Eq.~(\ref{kldiv}) cannot be performed analytically we evaluate it
numerically. 
We always pin the initial position of the tagged-particle at
$x_0=-2$. According to the effect of the pinning on the relaxation of
the tagged-particle, the scenario in which we tag the first
particle is referred to
as \emph{'confining'} (since background particles obstruct the
relaxation of the tagged-particle) and the one in which we tag the first particle as
\emph{'pushing'} (since background particles exert an
entropic force pushing the tagged-particle over the barrier).  $
\mathcal{D}(t,x_{0i})$ as a function of the barrier height $U_b$ for
$N=5$ and $N=9$ is shown in Fig.~\ref{fig:klrelax}.

Note that $\lim_{t\to 0}\mathcal{D}(t,x_{0i})=\infty$ irrespective of
$N$ and $U_b$ since we are comparing a delta distribution with a
smooth probability density.
Conversely, in an arbitrarily small time interval
$\tau_\epsilon>0$ the non-Markovian tagged-particle density $\mathcal{G}(x,t|x_{0i})$ evolves to a smooth,
well-behaved probability density such that $\mathcal{D}(t>0,x_{0i})$
is always finite and the 'pathology' at $t=0$ is mathematical and not physical.

With this in mind we observe in Fig.~\ref{fig:klrelax} a striking difference between the 'confining' and 'pushing'
scenario. In the 'confining' setting $\mathcal{D}(t,x_{01})$ at a fixed
time $t$ is a monotonically increasing function of $U_b$ and as a
function of time decays on a time-scale that seems to be rather
independent of $U_b$. In the 'confining' scenario an increase of
$U_b$ displaces the system at $t=0^+$  further from equilibrium. This
is intuitive because $\mathcal{P}_{\rm eq}(x_1)$ becomes more strongly
confined to the boundary and hence away from $x_0$. To a dominant
extent relaxation occurs already on time-scales $t\gtrsim 1\ll
\Lambda_{\mathbf{1}}^{-1}$. The reason may be found in the fact that
$\Lambda_{\mathbf{1}}^{-1}$ corresponds to the mixing/ergodic time-scale on
which the full single-file (and thus the tagged-particle) explores the
entire system. In the 'confining' scenario the background particles are
drawn from a distribution that resembles closely the unconstrained
equilibrium and, in addition, the tagged-particle is nominally
unlikely to be found in the right well in equilibrium. Therefore, the
fraction of paths that cross the barrier in the ensemble of relaxation
paths is small, rendering
$V_{\mathbf{0k}}V_{\mathbf{k0}}$ for low-lying $\mathbf{k}$
essentially negligible (see Eq.~(\ref{NMGreen})). Nevertheless, a
second, slower relaxation stage is still discernible at $t\gtrsim 1$.
 
Conversely, in the 'pushing' scenario depicted in
Fig.~\ref{fig:klrelax}b and \ref{fig:klrelax}d we find (i) the dependence of
$\mathcal{D}(0^+,x_{01})$ on $U_b$ to be inverted, and (ii) for
given $N$ and $U_b$ relaxation extends to much longer time-scales
compared to the 'confining' scenario.
In order to
rationalize (i) we consider a pair of barriers $U_{b_1},U_{b_2}$ and take the limit
\begin{equation}
    \lim_{t\to
      0}\left(\mathcal{D}^{b_1}(t,x_0)-\mathcal{D}^{b_2}(t,x_0)\right)=\ln\left(\mathcal{P}^{b_2}_{\rm
        eq}(x_0)/\mathcal{P}^{b_1}_{\rm eq}(x_0)\right),
\label{limit}
\end{equation}
which is finite and well defined despite the fact that
$\lim_{t\to 0}\mathcal{D}^{b_1,b_2}(t,x_0)$ are
infinite. Eq.~(\ref{limit}) explains that the dependence of
$\mathcal{D}(0^+,x_0)$ on $U_b$ is not unique and depends on
the pinning point $x_0$ which determines whether or not $\mathcal{P}^{b_2}_{\rm
        eq}(x_0)/\mathcal{P}^{b_1}_{\rm eq}(x_0)$ is greater or
smaller than $1$ (see Fig. ~\ref{fig:meanbarrier}b and \ref{fig:meanbarrier}d). (ii) can be
understood by an extension of the argument put forward in the
discussion of the 'confining' scenario, i.e. as a result of the
pinning the initial configurations of the background particles are
displaced much further away from equilibrium, rendering $V_{\mathbf{0k}}V_{\mathbf{k0}}$ for low-lying $\mathbf{k}$
substantial (see Eq.~(\ref{NMGreen})). Therefore, a pronounced second
relaxation stage is visible at longer times $t\gtrsim 1$. 

\begin{figure}
    \centering
    \includegraphics[width=0.48\textwidth]{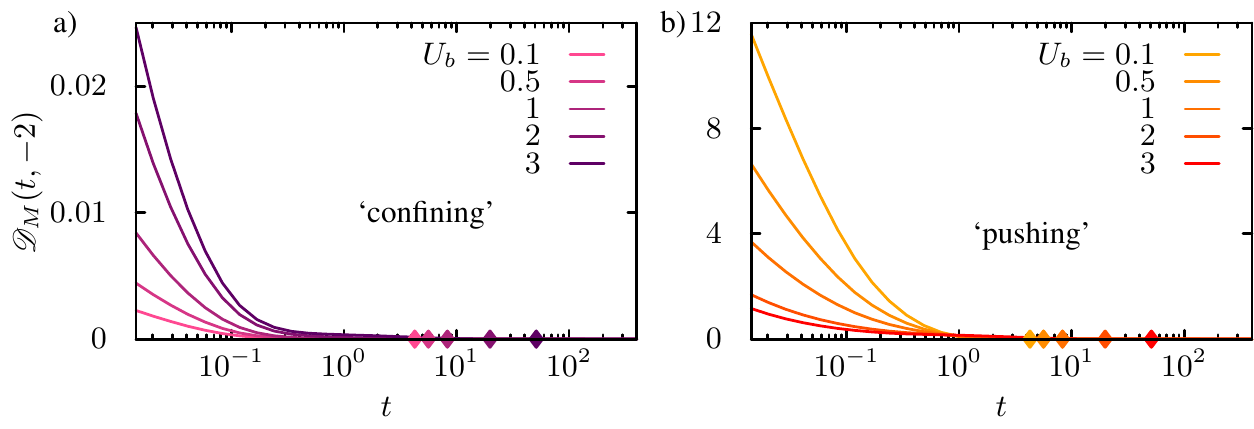}
    \caption{Time evolution of $\mathcal{D}_M(t,x_{0i})$ for various
      barrier-heights $U_b$ for $N=5$ (a -- confining, $i=1$; b -- pushing, $i=N$). The
      symbols denote $\mathcal{D}_M(\Lambda_{\mathbf{1}}^{-1},x_{0i})$.}
    \label{FMarkov}
\end{figure}

Based on Fig.~\ref{fig:klrelax} alone we are not able to deduce
whether these observations are a trivial consequence of the pinning
in the sense that they have nothing to do with memory (note that
a Markov process 'remembers' the initial condition up to $\sim
\tau_{\rm rel}$) or whether they are in fact a signature of memory in
the dynamics. Additional insight is gained by inspecting the
relaxation of the full, Markovian single-file evolving from the same
initial condition, i.e.
\begin{equation}
    \mathcal{D}_M(t,x_0)\equiv
    \left[\prod_{i=1}^N\int_{-\pi}^{\pi} dx_i\right]
    G(\mathbf{x},t,P_0)\ln\left(\frac{G(\mathbf{x},t,P_0)}{P_{\rm
      eq}(\mathbf{x})}\right),
    \label{kldivM}
\end{equation}
were we have introduced the joint Markovian two-point probability
density $G(\mathbf{x},t,P_{0})\equiv
\int d\mathbf{y}_0G(\mathbf{x},t|\mathbf{y}_{0})P_0(\mathbf{y}_0)$
whereby, for $-\pi<x_0<-\pi/2$, $P_0(y)$ is defined as
\begin{equation}
P_0(\mathbf{x})=N!(\pi+x_0)^{-N_L}\hat{O}_{\mathbf{x}}\delta(x_{i}-x_0)\frac{\mathrm{e}^{-U_b\sum_{j=i+1}^N\theta(\pi/2-|x_j|)}}{(\pi\mathrm{e}^{-U_b}-x_0)^{N_R}},
\label{init}
\end{equation}
where $\theta(x)$ is the Heaviside step function, and $P_{\rm
  eq}(\mathbf{x})=\lim_{t\to\infty}G(\mathbf{x},t|\mathbf{x}_{0})$. The
integration in Eq.~(\ref{kldivM}) can be performed analytically (for
details please see Ref.~\onlinecite{Lapolla_PRL}).
Introducing the two-point joint density of the single-particle problem
$\Gamma_t(x,a,b)\equiv\sum_{k}\psi^R_k(x)[\int_{a}^{b}dy\psi^L_k(y)P_{0}(y)]\mathrm{e}^{-\lambda_kt}$
with $P_{0}(y)\equiv \theta(-\pi/2-y)/(\pi+x_0)+\theta(y+\pi/2)\mathrm{e}^{-U(y)}/(\pi\mathrm{e}^{-U_b}-x_0)$
as well as the auxiliary function
\begin{equation}
\Xi_t(a,b)=\int_a^bdx\Gamma_t(x,a,b)\ln(\Gamma_t(x,a,b)/P_{\rm eq}(x)),
\label{aux2}
\end{equation}  
where $P_{\rm eq}(x)=\mathrm{e}^{-U(x)}/\pi(1+\mathrm{e}^{-U_b})$, the 
result reads
\begin{equation}
\mathcal{D}_M(t,x_{0})=\Xi_t(-\pi,\pi)\Xi_t(-\pi,x_0)^{N_L}\Xi_t(x_0,\pi)^{N_R}.
\label{KLM}
\end{equation}
An explicit solution is obtained with the aid of \textsf{Mathematica}
\cite{Mathematica}. As it is bulky but straightforward we do not show
it here.
The Markovian result in Eq.~(\ref{KLM}) for the same set of parameters
as in Fig.~\ref{fig:klrelax}a and \ref{fig:klrelax}b is depicted in Fig.~\ref{FMarkov}.
A comparison of Figs.~\ref{fig:klrelax} and  \ref{FMarkov} reveals
that the second, long-time relaxation stage observed in the 'pushing'
scenario in Fig.~\ref{fig:klrelax} is absent in the Markovian
setting (compare Figs.~\ref{fig:klrelax} and  \ref{FMarkov} and note
that the relaxation time $\Lambda_{\mathbf{1}}^{-1}$ is identical in
both settings).  This
in turn implies that the pronounced second
relaxation stage in the non-Markovian, tagged-particle scenario at times
$t\gtrsim 1$ is indeed a signature of memory.

\section{Discussion}
We identified pronounced signatures of memory in the overdamped relaxation of a
tagged-particle in a single-file confined to a bi-stable potential. On
the level of linear correlations in equilibrium memory is visible in
the form of a multi-scale relaxation of the autocorrelation function
(see Fig.~\ref{fig:autocorr}),  \textcolor{black}{a substantial and
  long-ranged (linear) memory-kernel (see Fig.~\ref{memKer}b)}, and
a seemingly paradoxical shortening of the so-called correlation
time $T_c$ (see Fig.~\ref{fig:corrtime}) . The latter was shown to be
an artifact of the definition of
$T_c$. When including the complete correlation-structure as encoded in the
so-called excess instantaneous free energy (see Eq.~(\ref{kldiv}))
distinctive signatures of memory emerge in the form of a second,
late-time relaxation regime.

The memory originates from the fact that the entire single-file
relaxes to equilibrium in the form of
linearly independent many-body eigenmodes, which become projected on the motion of a tagged
particle
\cite{lapolla_unfolding_2018,lapolla_manifestations_2019}. The
projection couples distinct modes thus braking Markovianity and giving
rise to memory \cite{lapolla_manifestations_2019}. It turns out to be
very important which particle is tagged. Here, we were only interested
in the 'confining' (all background particles in front of the tagged
particle) and  'pushing' (all background particles behind the tagged
particle) scenarios and found qualitatively different relaxation
behavior. A systematic analysis would be required to understand
the intricate details in how the number of particles on each side
affects relaxation dynamics, which is beyond the scope of
\textcolor{black}{the present work}.

\textcolor{black}{We have shown that the memory non-trivially depends
  on the external potential. That is, a tagged particle was shown to experience
  the external potential directly (i.e. time-locally) as well as
  indirectly through the effect it exerts on the memory kernel. This
  effect is general  -- it occurs whenever the potential is
  sufficiently strong and either also acts on
  the latent degrees of freedom (i.e. those that are integrated out)
  or the interaction between the tagged-particle and the latent
  degrees of freedom is \textcolor{black}{not harmonic or, more
    generally,} non-negligible. Direct evidence for the effect has
  been found e.g. in all-atom computer simulations
  of the hydration of molecular solutes
  \cite{Netz_PRX}.  It is important to keep this in mind when
  applying generalized Langevin
  equations (GLEs) with phenomenological memory kernels or microscopically consistent GLEs
  ``decorated''  with an external potential, as these may lead to
  erroneous conclusions or misinterpretations.}

\textcolor{black}{The
  application of the methodology for extracting and analyzing memory
  in tagged-particle dynamics 
  put forward in Eqs.~(\ref{autocor}), (\ref{kernel}), and
  ~(\ref{kldiv}) does \emph{not} require $C_i(t)$ nor
  $\mathcal{G}(x,t|x_{0i})$ to be known analytically. The quantities
  can equally well be determined from experiments or computer
  simulations. The analysis is expected to provide insight into memory
  effects as long as either the ``crowding'' (i.e. the concentration
  of background particles) and/or the barrier-height can be
  controlled. The qualitative features of the signatures of memory
  are expected to be preserved in most systems of effectively
  one-dimensional systems with obstructed tagged-particle
  dynamics. The proposed analysis of memory effects can be viewed as complementary
  to the analysis of anomalies in tagged-particle diffusion \cite{toolbox,Schwarzl,Metzler_2014}.} 

Our results can readily be tested by existing experiments probing colloidal
particle systems (see e.g. Ref.~\onlinecite{Wei625,Hanes_2012,Thorneywork_2020}),
and may furthermore be relevant for a theoretical description of transport in ion-channels \cite{roux_theoretical_2004,
  pohorille_validity_2017, kopec_direct_2018,
  epsztein_towards_2020}. Our results can be extended in diverse ways,
most immediately by including other types of inter-particle
interactions \cite{Kollman} and time-depended energy barriers \cite{subrt_diffusion_2006}.

\section{Appendix}
In this Appendix we give explicit expressions for the single-particle
eigenfunctions that are required in the diagonalization of the
many-body Fokker-Planck operator using the so-called coordinate Bethe
ansatz. For
details of the solution method please see
\cite{lapolla_unfolding_2018,lapolla_manifestations_2019,lapolla_bethesf_2020}). 
The eigenfunctions of the corresponding single-particle eigenvalue
problem 
\begin{eqnarray}
(\partial_{x}^2+\partial_{x}\{\partial_{x}U(x)\})\psi_{k}^{R}(x)&=&-\lambda_k\psi_{k}^{R}(x)\nonumber\\
  (\partial_{x}^2-\{\partial_{x}U(x)\}\partial_{x})\psi_{k}^{L}(x)&=&-\lambda_k\psi_{k}^{L}(x),
\end{eqnarray}
where $\psi_{k}^{L,R}(x): [-\pi,\pi]\to\mathbb{R}$
allow for a spectral decomposition of the single-particle Green's
function
\begin{equation}
\Gamma(x,t|x_0)=\sum_k\psi_{k}^{R}(x)\psi_{k}^{L}(x_0)\mathrm{e}^{-\lambda_kt}.
\end{equation}  
 $\psi_{k}^{L,R}(x)$ enter Eq.~(\ref{eigf}) and are here defined via their 'Hermitianized'
counterpart $\psi_k(x):[-\pi,\pi]\to\mathbb{R}$ as
$\psi_k^R(x)=\ee{-U(x)/2}\psi_k(x)$,
$\psi_k^L(x)=\ee{U(x)/2}\psi_k(x)$ where\footnote{We here correct typos found in
  the original publications.}
\begin{align}
    &\psi_k(x)=\sqrt{\frac{2}{1+\delta_{k,0}}}\frac{\ee{-U(x)/2}}{\sqrt{\pi(1+\ee{-f0})}}\cos(\sqrt{\lambda_k}x), \quad \mathrm{mod}(k,4)=0\\
    &\psi_k(x)=\begin{dcases}
        -\frac{\cos(\sqrt{\lambda_k}(x+\pi))}{\sqrt{\pi}},& x<-\pi/2 \\
        \frac{\sin(\sqrt{\lambda_k}x)}{\sqrt{\pi}},& |x|\leq \pi/2 \\
        \frac{\cos(\sqrt{\lambda_k}(x-\pi))}{\sqrt{\pi}},& x>\pi/2
    \end{dcases}, \quad \mathrm{mod}(k,4)=1\\
     &\psi_k(x)=\sqrt{2}\frac{\ee{U(x)/2}}{\sqrt{\pi(1+\ee{f0})}}\cos(\sqrt{\lambda_k}x), \quad \mathrm{mod}(k,4)=2\\
    &\psi_k(x)=\begin{dcases}
        \frac{\cos(\sqrt{\lambda_k}(x+\pi))}{\sqrt{\pi}},& x<-\pi/2 \\
        \frac{\sin(\sqrt{\lambda_k}x)}{\sqrt{\pi}},& |x|\leq \pi/2 \\
        -\frac{\cos(\sqrt{\lambda_k}(x-\pi))}{\sqrt{\pi}},& x>\pi/2
    \end{dcases}, \quad \mathrm{mod}(k,4)=3,
\end{align}
where $\delta_{k,0}$ is the Kronecker delta.

\section*{Supplementary Material}
The Supplementary Material contains and extension of the code published in
Ref.~\onlinecite{lapolla_bethesf_2020} that
implements the analytical results presented this manuscript.


%
%

%

\begin{acknowledgments}
 The financial support from the German Research Foundation (DFG)
 through the Emmy Noether Program GO 2762/1-1 (to AG) is gratefully
 acknowledged. We thank Kristian Blom for fruitful discussions.
\end{acknowledgments}

\section*{DATA AVAILABILITY}
The data that support the findings of this study are available from
the corresponding author upon reasonable request.

\section*{References}
\bibliographystyle{unsrt}
\bibliography{Bistable}

\end{document}